\documentstyle[12pt]{article}

\def\0{\over } \def\1{\vec } \def\2{{1\over2}} \def\4{{1\over4}}
\def\5{\bar }
\def\6{\partial }

\let\a=\alpha \let\b=\beta \let\g=\gamma \let\d=\delta
\let\e=\epsilon

\def\({\left(} \def\){\right)} \def\<{\langle } \def\>{\rangle }
\def\[{\left[} \def\]{\right]}

\def\k{{\bf k}}

\def\be{\begin{equation}}
\def\ee{\end{equation}}
\def\bea{\begin{eqnarray}}
\def\eea{\end{eqnarray}}

\catcode`\@=11

\def\ps@myheadings{\let\@mkboth\@gobbletwo
\def\@oddhead{\hbox{} 
\rightmark\hfil\ninerm\thepage}
\def\@oddfoot{}\def\@evenhead{\ninerm\thepage\hfil 
\leftmark\hbox{}}\def\@evenfoot{}
\def\sectionmark##1{}\def\subsectionmark##1{}}

\evensidemargin
0.30truein
\begin{document}
\begin{titlepage}
\begin{flushright}
{\tt hep-ph/9310202}\\
{BI-TP 93/52}\\
\end{flushright}
\vfill
\begin{center}
{\Large GAUGE-INDEPENDENT EXTRACTION OF\\
THE NEXT-TO-LEADING-ORDER DEBYE MASS\\
FROM THE GLUON PROPAGATOR\\}
\vfill
{\large A. K. Rebhan}\footnote{~~On leave of absence from
Institut f\"ur Theoretische Physik der Technischen Universit\"at Wien,
A--1040 Vienna, Austria.
Address after November 1, 1993: DESY, Gruppe Theorie, D--22603 Hamburg,
Germany.}\\
\bigskip
{\it Fakult\"at f\"ur Physik, Universit\"at Bielefeld\\
D-33501 Bielefeld, Germany}\\
\vfill
{\large ABSTRACT}
\end{center}
\begin{quotation}
\noindent It is shown that by defining the Debye mass through
the relevant pole of the static gluon propagator
rather than the zero-momentum limit of the
time-time component of the gluon self-energy, a gauge-independent
result for the next-to-leading order correction can be derived
upon resummation of hard-thermal-loop contributions.
The result turns out to be logarithmically
sensitive to the magnetic screening mass.

\noindent{\sl
[Talk given at the 3rd Workshop on Thermal Field Theories and
their Applications, August 15--27, 1993, Banff, Alberta, Canada]}
\end{quotation}
\vfill
\begin{flushleft}
October 1993
\end{flushleft}
\end{titlepage}

\newcommand{\symbolfootnote}{\renewcommand{\thefootnote}
        {\fnsymbol{footnote}}}
\renewcommand{\thefootnote}{\fnsymbol{footnote}}
\newcommand{\alphfootnote}
        {\setcounter{footnote}{0}
         \renewcommand{\thefootnote}{\sevenrm\alph{footnote}}}

\newcounter{sectionc}\newcounter{subsectionc}\newcounter{subsubsectionc}
\renewcommand{\section}[1] {\vspace{0.6cm}\addtocounter{sectionc}{1}
\setcounter{subsectionc}{0}\setcounter{subsubsectionc}{0}\noindent
        {\bf\thesectionc. #1}\par\vspace{0.4cm}}
\renewcommand{\subsection}[1] {\vspace{0.6cm}\addtocounter{subsectionc}{1}
        \setcounter{subsubsectionc}{0}\noindent
        {\it\thesectionc.\thesubsectionc. #1}\par\vspace{0.4cm}}
\renewcommand{\subsubsection}[1]
{\vspace{0.6cm}\addtocounter{subsubsectionc}{1}
        \noindent {\rm\thesectionc.\thesubsectionc.\thesubsubsectionc.
        #1}\par\vspace{0.4cm}}
\newcommand{\nonumsection}[1] {\vspace{0.6cm}\noindent{\bf #1}
        \par\vspace{0.4cm}}

\newcounter{appendixc}
\newcounter{subappendixc}[appendixc]
\newcounter{subsubappendixc}[subappendixc]
\renewcommand{\thesubappendixc}{\Alph{appendixc}.\arabic{subappendixc}}
\renewcommand{\thesubsubappendixc}
        {\Alph{appendixc}.\arabic{subappendixc}.\arabic{subsubappendixc}}

\renewcommand{\appendix}[1] {\vspace{0.6cm}
        \refstepcounter{appendixc}
        \setcounter{figure}{0}
        \setcounter{table}{0}
        \setcounter{equation}{0}
        \renewcommand{\thefigure}{\Alph{appendixc}.\arabic{figure}}
        \renewcommand{\thetable}{\Alph{appendixc}.\arabic{table}}
        \renewcommand{\theappendixc}{\Alph{appendixc}}
        \renewcommand{\theequation}{\Alph{appendixc}.\arabic{equation}}
        \noindent{\bf Appendix \theappendixc #1}\par\vspace{0.4cm}}
\newcommand{\subappendix}[1] {\vspace{0.6cm}
        \refstepcounter{subappendixc}
        \noindent{\bf Appendix \thesubappendixc. #1}\par\vspace{0.4cm}}
\newcommand{\subsubappendix}[1] {\vspace{0.6cm}
        \refstepcounter{subsubappendixc}
        \noindent{\it Appendix \thesubsubappendixc. #1}
        \par\vspace{0.4cm}}

\def\abstracts#1{{
        \centering{\begin{minipage}{30pc}\tenrm\baselineskip=12pt\noindent
        \centerline{\tenrm ABSTRACT}\vspace{0.3cm}
        \parindent=0pt #1
        \end{minipage} }\par}}

\newcommand{\bibit}{\it}
\newcommand{\bibbf}{\bf}
\renewenvironment{thebibliography}[1]
        {\begin{list}{\arabic{enumi}.}
        {\usecounter{enumi}\setlength{\parsep}{0pt}
\setlength{\leftmargin 1.25cm}{\rightmargin 0pt}
         \setlength{\itemsep}{0pt} \settowidth
        {\labelwidth}{#1.}\sloppy}}{\end{list}}

\topsep=0in\parsep=0in\itemsep=0in
\parindent=1.5pc

\newcounter{itemlistc}
\newcounter{romanlistc}
\newcounter{alphlistc}
\newcounter{arabiclistc}
\newenvironment{itemlist}
        {\setcounter{itemlistc}{0}
         \begin{list}{$\bullet$}
        {\usecounter{itemlistc}
         \setlength{\parsep}{0pt}
         \setlength{\itemsep}{0pt}}}{\end{list}}

\newenvironment{romanlist}
        {\setcounter{romanlistc}{0}
         \begin{list}{$($\roman{romanlistc}$)$}
        {\usecounter{romanlistc}
         \setlength{\parsep}{0pt}
         \setlength{\itemsep}{0pt}}}{\end{list}}

\newenvironment{alphlist}
        {\setcounter{alphlistc}{0}
         \begin{list}{$($\alph{alphlistc}$)$}
        {\usecounter{alphlistc}
         \setlength{\parsep}{0pt}
         \setlength{\itemsep}{0pt}}}{\end{list}}

\newenvironment{arabiclist}
        {\setcounter{arabiclistc}{0}
         \begin{list}{\arabic{arabiclistc}}
        {\usecounter{arabiclistc}
         \setlength{\parsep}{0pt}
         \setlength{\itemsep}{0pt}}}{\end{list}}

\newcommand{\fcaption}[1]{
        \refstepcounter{figure}
        \setbox\@tempboxa = \hbox{\tenrm Fig.~\thefigure. #1}
        \ifdim \wd\@tempboxa > 6in
           {\begin{center}
        \parbox{6in}{\tenrm\baselineskip=12pt Fig.~\thefigure. #1 }
            \end{center}}
        \else
             {\begin{center}
             {\tenrm Fig.~\thefigure. #1}
              \end{center}}
        \fi}

\newcommand{\tcaption}[1]{
        \refstepcounter{table}
        \setbox\@tempboxa = \hbox{\tenrm Table~\thetable. #1}
        \ifdim \wd\@tempboxa > 6in
           {\begin{center}
        \parbox{6in}{\tenrm\baselineskip=12pt Table~\thetable. #1 }
            \end{center}}
        \else
             {\begin{center}
             {\tenrm Table~\thetable. #1}
              \end{center}}
        \fi}

\def\@citex[#1]#2{\if@filesw\immediate\write\@auxout
        {\string\citation{#2}}\fi
\def\@citea{}\@cite{\@for\@citeb:=#2\do
        {\@citea\def\@citea{,}\@ifundefined
        {b@\@citeb}{{\bf ?}\@warning
        {Citation `\@citeb' on page \thepage \space undefined}}
        {\csname b@\@citeb\endcsname}}}{#1}}

\newif\if@cghi
\def\cite{\@cghitrue\@ifnextchar [{\@tempswatrue
        \@citex}{\@tempswafalse\@citex[]}}
\def\citelow{\@cghifalse\@ifnextchar [{\@tempswatrue
        \@citex}{\@tempswafalse\@citex[]}}
\def\@cite#1#2{{$\null^{#1}$\if@tempswa\typeout
        {IJCGA warning: optional citation argument
        ignored: `#2'} \fi}}
\newcommand{\citeup}{\cite}

\def\fnm#1{$^{\mbox{\scriptsize #1}}$}
\def\fnt#1#2{\footnotetext{\kern-.3em
        {$^{\mbox{\sevenrm #1}}$}{#2}}}

\font\twelvebf=cmbx10 scaled\magstep 1
\font\twelverm=cmr10 scaled\magstep 1
\font\twelveit=cmti10 scaled\magstep 1
\font\elevenbfit=cmbxti10 scaled\magstephalf
\font\elevenbf=cmbx10 scaled\magstephalf
\font\elevenrm=cmr10 scaled\magstephalf
\font\elevenit=cmti10 scaled\magstephalf
\font\bfit=cmbxti10
\font\tenbf=cmbx10
\font\tenrm=cmr10
\font\tenit=cmti10
\font\ninebf=cmbx9
\font\ninerm=cmr9
\font\nineit=cmti9
\font\eightbf=cmbx8
\font\eightrm=cmr8
\font\eightit=cmti8


\centerline{\tenbf GAUGE-INDEPENDENT EXTRACTION OF}
\baselineskip=16pt
\centerline{\tenbf THE NEXT-TO-LEADING-ORDER DEBYE MASS}
\centerline{\tenbf FROM THE GLUON PROPAGATOR}
\vspace{0.8cm}
\centerline{\tenrm A. K. REBHAN}
\baselineskip=13pt
\centerline{\tenit Fakult\"at f\"ur Physik, Universit\"at Bielefeld}
\baselineskip=12pt
\centerline{\tenit D-33501 Bielefeld, Germany}
\vspace{0.9cm}
\abstracts{It is shown that by defining the Debye mass through
the relevant pole of the static gluon propagator
rather than the zero-momentum limit of the
time-time component of the gluon self-energy, a gauge-independent
result for the next-to-leading order correction can be derived
upon resummation of hard-thermal-loop contributions.
The result turns out to be logarithmically
sensitive to the magnetic screening mass.}

\vspace{0.8cm}
\twelverm   
\baselineskip=14pt
\section{The Hot QCD Debye Mass Puzzle}

In QED, the electric permittivity $\e(\omega,k)$ is
given by the time-time component of the photon self-energy,
\be
\e(\omega,k)=1+{\Pi_{00}(\omega,k)\0\k^2}.
\label{eps}\ee
In the static limit $\omega=0$, this is the screening factor
of longitudinal electric fields\cite{Kapusta},
\be
\<E^i_L(k)\>=-i{k^i J^0 \0 \k^2+\Pi_{00}(0,k)}
\label{elong}\ee
where $J^\mu$ is a weak, conserved external current.
The long-wavelength limit of $\Pi_{00}(0,k)$ is
usually identified with the Debye screening mass,
$m_{el.}^2 \stackrel{\rm def}= \Pi_{00}(0,k\to0)$.

In fact, in the high-temperature limit ($T\gg k$) the
leading contribution
\be
\Pi_{00}(0,k)={e^2T^2\03}
\label{m0}\ee
provides a momentum-independent mass term, which
modifies the classical Coulomb potential by a factor
$e^{-m_{\rm el.}r}$.
The leading high-temperature result in
QCD is very similar; one just has to replace
$e^2 \to g^2(N+N_f/2)$ for color group SU($N$) and $N_f$
fermions\cite{Kapusta}.

In QED, higher-order corrections to
the long-wavelength limit of $\Pi_{00}(0,k)$ are known
through an exact relation\cite{Fradkin} to the equation of state
\be
\Pi_{00}(0,k\to0)=e^2{\6^2 P(\mu,T)\0\d\mu^2},
\label{eos}\ee
where $P(\mu,T)$ is the pressure and $\mu$ the chemical potential.
Since the first three terms in the perturbation expansion of $P$
are known, one also knows the first three terms in (\ref{eos}).
With $\mu=0$,\cite{Kapusta}
\be
\Pi_{00}(0,k\to0)={e^2T^2\03}\(1-{3e^2\08\pi^2}+{\sqrt3 e^3\04\pi^3}
+\ldots\).
\label{pi00qed}
\ee

In QCD, one would expect the first correction term to be of
relative order $g$ rather than $g^2$, because of the ``plasmon
effect''\cite{Kapusta}. However, after ring resummation
a gauge-fixing dependent result\cite{T}
is found for $\Pi_{00}(0,k\to0)$ in covariant gauges
\be
\Pi_{00}(0,k\to0)=m^2_{0\rm el.}\(
1+\a {N\04\pi}\sqrt{6\02N+N_f} g \)
\label{toi}\ee
with gauge fixing parameter $\a$,
which shows that it cannot be identified with a directly
measurable quantity that the Debye mass presumably should be.

The reason for the gauge dependence of (\ref{toi}) of course
is that the self-energy of non-Abelian gauge fields is a
gauge variant quantity. In QED, on the other hand, the self-energy
of photons is directly related to the correlator of the
gauge-invariant electromagnetic
field strengths, so this problem does not arise there.

It was therefore argued\cite{KK1,KK2,Kapusta} that the
non-Abelian Debye mass cannot in general
be derived from the gluon propagator except in the
temporal axial gauge, where the time-time component of the
propagator is again directly related to the correlator
of two chromoelectric field operators.

The temporal axial gauge, however, is notoriously difficult
at finite temperature\cite{PVL}, and actually the principal-value
prescription commonly used to deal with its singularities
at zero frequencies has proved to be flawed\cite{CCM}. These singularities
moreover prevent a straightforward implementation of
ring resummation, because the static mode cannot be isolated.
A first attempt\cite{KK1} of a ring-resummed calculation of the non-Abelian
Debye mass yielded a negative correction term
to the Debye mass
of relative order $g$, but this could not be reproduced by taking the
temporal limit of a corresponding calculation in general axial
gauge\cite{FK}, which by the way gave a positive result. The former
was consequently withdrawn and
replaced\cite{KK2} by a calculation
which resums the asymptotic gluon mass rather than the
leading-order Debye mass (not taking into account
vertex corrections, however). This yielded again a negative
correction to the Debye mass as defined through the
infrared limit of the self-energy in temporal axial gauge, to wit,
\be
\Pi_{00}^{\rm TAG}(0,k\to0)\approx
m^2_{0\rm el.}\(1-{g\02\pi}\sqrt{{3\02}(N+N_f/2)}\).
\label{mkk}\ee

Leaving aside for the moment the open questions about pole prescriptions in
temporal axial gauge and even which resummation scheme should be
employed, there still remains the question whether the analysis
of correlators of chromoelectric field operators really guarantees to
give gauge independent answers where the gluon propagator evidently
failed. The temporal axial gauge is singled out as the one where
no higher vertex functions are needed to obtain this correlator,
but one could of course use any gauge. Because the chromoelectric
field is not a gauge-invariant operator, there is a priori no
reason to expect gauge fixing independence of its correlation functions.
Indeed, one finds that under a change of gauge condition $f_\mu A^\mu\to
(f_\mu+\d f_\mu)A^\mu$ the correlator of two chromoelectric field
operators varies according to\cite{KKRS}
\be
\delta\<E^a_j(x)E^e_k(y)\>=-gf^{abc}\int d^4\!z
\< E^b_j(x)\5c^c(x)c^d(z)\d f_\mu A^{d\mu}(z)E^e_k(y)\>
+(a,j,x \leftrightarrow e,k,y),
\ee
where $\5c$ and $c$ are Faddeev-Popov ghost fields, which make
their appearance even in gauges which are otherwise ghost-free.

Thus one is not necessarily on safe grounds by studying
the correlation of field operators rather than of the gauge fields.
On the other hand, if one succeeded in extracting gauge independent
information from the gauge variant quantity $\Pi_{00}$, then
this information should equally be found in the correlator of
electric field operators, since in a particular gauge (the temporal one)
the two are directly related.

\section{Changing the Definition of the Debye Mass}

A strong hint that the
very definition of the Debye mass as $\Pi_{00}(0,k\to0)$
might not be sufficient beyond leading order is seen by the consequences
of keeping the term proportional to $\k^2$ in the resummed result
(\ref{toi}). Then\cite{N1}
\be
\Pi_{00}(0,k\to0)=m^2_{0\rm el.}\(
1+\a {N\04\pi}\sqrt{6\02N+N_f} g \) - {2N\03\pi}\sqrt{6\02N+N_f} g\k^2
+\ldots,
\label{nad}\ee
which, when inserted into (\ref{elong}) leads to a different
mass term besides an over-all factor that is constant.
This does not remove the gauge dependence of the putative correction
to the Debye mass (it just replaces $\a$ by $\a+{8\03}$)%
\cite{N1}, but it does
tell that the $k$-dependence of $\Pi_{00}(0,k)$ still can change
things!

Let us therefore go back to the linear response formula for
longitudinal electric fields, eq.~(\ref{elong}). Actually, this
formula is valid\cite{Weldon} also in the non-Abelian case, since
with a single source $J$ there is also only a single direction
in color space to which the gauge potentials can point, and
the nonlinear terms in the chromoelectric field strength vanish
trivially. If $J_0$ is a static point charge $Q$
located at the origin, then the Fourier
transform of eq.~(\ref{elong}) equals minus the gradient of
a potential $\Phi$  given by
\begin{eqnarray}
&&\Phi(r)= Q\int\frac{d^3k}{(2\pi)^3}
\frac{e^{i{\bf k}{\bf r}}}{k^2+\Pi_{00}(k_0=0,k)}\nonumber\\
&=&\frac{Q}{(2\pi)^2}\int_{-\infty}^\infty
\frac{e^{ikr}-e^{-ikr}}{2ir} \frac{k\,dk}{k^2+\Pi_{00}(0,k)}.
\label{v}\end{eqnarray}

The usual definition of the Debye mass as
$m_{el.}^2 \stackrel{\rm def}= \Pi_{00}(0,k\to0)$
is commonly motivated by saying that when $r$ is very large,
the dominant contribution to the integral
(\ref{v}) comes from $k=0$. This is not
quite correct, however. Inserting for instance the
leading high-temperature result (\ref{m0}), which is
a simple constant mass squared, one can readily evaluate
(\ref{v}) by appropriately closing the contour in the complex
$k$-plane wherein the integrand has simple poles at $k=\pm im_{0\rm el.}$,
where only ${\rm Re}\,k=0$. If $\Pi_{00}(0,k)$ depends on $k$,
as it will be the case in general
(i.e., beyond leading order), then this dependence will
be important to determine the location of the pole and thus
the magnitude of the mass that will appear in the exponent of
$e^{-mr}$. It is therefore rather obvious that one should
define the Debye mass, which certainly should account for the actual
exponential fall-off, self-consistently by the zeros of
the denominator in (\ref{v}). I therefore propose the implicit
definition
\be
m^2_{\rm el.} = \Pi_{00}(0,k)\Big|_{k^2=-m^2_{\rm el.}}
\label{mdef}\ee
for the (chromo)electric Debye mass.

The Debye mass is thus defined through the singularity of the
propagator appearing in (\ref{v}), and this propagator is in
fact one of the components of the (static) gauge-field propagator
for which formal arguments showing gauge independence of its
poles can be derived from gauge fixing identities\cite{KKR}.
In non-Abelian gauge theories, the gauge dependence of the
usual definition of the Debye mass through the zero-momentum
limit of the self-energy is only to be expected
because the leading high-temperature corrections
move the pole away from strictly $k=0$.

One might now wonder why there appeared to be no problem with
the old definition in the case of QED. In QED there is no
problem of gauge dependence for the photon self-energy, for
the latter is gauge independent through all orders of perturbation
theory. Nevertheless, if one wants the Debye mass to truly
describe the exponential fall-off of the electrostatic
potential, then it is clear that one has to adopt the self-consistent
definition (\ref{mdef}). But then there is a correction term to
be added to (\ref{pi00qed}) according to
\be
m^2_{\rm el.}=\Pi_{00}(0,k\to0)+\[\Pi_{00}(0,k)\big|_{k^2=-m^2_{\rm el.}}
-\Pi_{00}(0,k\to0)\],
\label{mqed}\ee
and this correction is
\be
m^2_{\rm el.}-\Pi_{00}(0,k\to0)=
m^2_{0\rm el.}\({2e^2\09\pi^2}-{e^2\06\pi^2}\[\ln{\tilde\mu\0\pi T}+\g_E\]
+O(e^4)\),
\label{dmqed}\ee
where $\tilde\mu$
is the mass scale introduced by dimensional regularization
in which minimal subtraction has been performed.
This correction term shows what has been missing in the first place
when trying to identify (\ref{pi00qed}) with a physical quantity.
As a physical quantity,
it has to be a renormalization-group invariant, which is
indeed what (\ref{dmqed}) brings about: the coefficient of the
logarithmic term in (\ref{dmqed}) is exactly such that
$\6m^2_{\rm el.}/\6\tilde\mu=0$
because $de/d(\ln\tilde\mu)=\b(e)=e^3/(12\pi^2)+O(e^5)$.

In fact, the importance of the $k$-dependence of $\Pi_{00}(0,k)$
for Debye screening has been recognized previously in the case of
a degenerate electron gas, giving rise to the so-called Friedel
oscillations\cite{FW}. Yet the self-consistent definition (\ref{mdef})
of the Debye mass has not been adopted before as far as I know.

\section{Next-to-Leading Order Calculation of the Non-Abelian Debye Mass}

A complete calculation of perturbative corrections to the 
high-temperature dispersion laws beyond those determined by the
gauge-independent hard thermal loops (HTL's) 
has been shown
by Braaten and Pisarski\cite{BP}
to require the resummation of all of the HTL contributions
to self-energy and vertices. 
This rather involved resummation scheme has been applied in particular
to determine the damping of collective excitations\cite{Damping}, and
most recently also to compute the next-to-leading order contribution
to the chromodynamical plasma frequency\cite{HS}.

The problem of calculating the next-to-leading order contribution to
the non-Abelian Debye mass is in fact a problem of the same kind.
With the definition (\ref{mdef}), the Debye mass can be regarded
as being given by the position of the plasmon pole when the
frequency is lowered below the plasma frequency and eventually put
to zero, whilst the wavevector becomes imaginary.

Fortunately, with
zero external frequencies the Braaten-Pisarski scheme can be simplified
tremendously\cite{AE}. In the imaginary time formalism, one may
separate the modes of
a resummed bosonic propagator $1/[(2\pi n T)^2+k^2+\Pi]$
into the static mode $n=0$ and the non-static ones, where the latter
are automatically hard in the terminology of Braaten and Pisarski.
Hence, only the static mode needs resummation of the HTL,
which in the propagator is just the leading-order Debye mass
$m_{0\rm el.}$.
This separation of course makes analytic continuation to non-zero
external frequencies rather impossible, but this is no shortcoming
in the static case. With all the external frequencies
being zero, all potentially soft lines of a diagram are static, and
so are the vertices that would need resummation of HTL
contributions. However, the latter vanish entirely in the purely
static limit.

Hence,
as far as the relative order $g$ correction to the static
gluon self-energy is concerned, 
Braaten-Pisarski resummation boils down to a conventionally
ring-resummed\cite{Kapusta} one-loop calculation.
The static ring-resummed propagator in general covariant gauge reads
\begin{equation}
\Delta_{\mu\nu}\bigg|_{p_0=0}=\left[
\frac1{{\bf p}^2+m_0^2}
\delta^0_\mu \delta^0_\nu
+\frac1{{\bf p}^2}
\left(\eta_{\mu\nu}-\delta^0_\mu \delta^0_\nu +
\frac{P_\mu P_\nu}{{\bf p}^2} \right)
+\alpha \frac{P_\mu P_\nu}{({\bf p}^2)^2} \right]_{P_0=0},
\label{stpr}\end{equation}
and the complete next-to-leading order contribution to $\Pi_{00}(0,k)$
is found as\cite{I}
\begin{eqnarray}
&&\delta\Pi_{00}(k_0=0,{\bf k})
=\underbrace{gmN\sqrt{\frac6{2N+N_f}}}_{g^2T}\int
\frac{d^{3-2\varepsilon}p}{(2\pi)^{3-2\varepsilon}}
\biggl\{\frac1{{\bf p}^2+m^2}+\frac1{{\bf p}^2} \nonumber\\
&&\qquad+\frac{4m^2-({\bf k}^2+m^2)
[3+2{\bf p}{\bf k}/{\bf p}^2]}{{\bf p}^2({\bf q}^2+m^2)}
+\alpha ({\bf k}^2+m^2)
\frac{{\bf p}^2+2{\bf p}{\bf k}}{{\bf p}^4({\bf q}^2+m^2)} \biggr\},
\label{pi00}\end{eqnarray}
where $m\equiv m_{0\rm el.}$ and
${\bf q}={\bf p}+{\bf k}$. (Here dimensional regularization has
been used when separating the static modes from the sum over
Matsubara frequencies\cite{AE}; the limit $\varepsilon\to0$ gives a
regular expression because of the odd integration dimension.)

The new definition (\ref{mdef}) requires to evaluate at $\k^2=-m^2$.
There the gauge parameter ($\a$) dependent part vanishes
algebraically, in accordance with the gauge fixing identities\cite{KKR}.
However, closer inspection reveals that the integrals in (\ref{pi00})
develop ``mass-shell'' singularities, caused by massless transverse
and massless unphysical modes in the gluon propagator.

The third term of the integrand in (\ref{pi00}) is logarithmically
singular as $k^2\to-m^2$, and the singularity is caused exclusively
by the massless denominator in the spatially transverse part of
the gluon propagator (\ref{stpr}). A magnetic screening mass
would remove this singularity, and because the latter is only
logarithmic, the coefficient of the corresponding logarithm is
determined by (\ref{pi00}),
\be
\d\Pi_{00}(0,k)\Big|_{k^2\to-m^2_{\rm el.}}\to
{g^2 N m_{\rm el.} T\02\pi}\ln{2m_{\rm el.}\0m_{\rm magn.}}
\label{pi00sing}\ee
up to terms that are regular as $m_{\rm magn.}\to0$.
Assuming that $m_{\rm magn.}\sim gm_{el.}$, the next-to-leading
order contribution to $m_{\rm el.}^2$ is found to be of order
$g\ln(1/g)$ rather than $g$,
\be
{\d m_{\rm el.}^2\0m_{0\rm el.}^2}=
\frac{N}{2\pi}\sqrt{\frac6{2N+N_f}}\,g\ln\frac1g+O(g),
\label{dln}\ee
and it is positive, at least at weak coupling $g\ll1$. Taken
seriously for larger coupling $g\sim1$, the logarithm would
eventually switch sign, but there the sublogarithmic terms
would be of equal importance. 

Unfortunately,
the sublogarithmic terms cannot be calculated completely, because
the presumed phenomenon of magnetic screening is
nonperturbative\cite{Linde}. However, in order to obtain an {\it
estimate} of those, let us assume that a simple
replacement of $1/\k^2\to1/(\k^2+m^2_{\rm magn.})$
in the transverse part of the static propagator (\ref{stpr})
correctly summarizes the effects at $k\sim g^2T$. Then we
may go on to evaluate the remaining contributions
in (\ref{pi00}).
Here one encounters a difficulty with the $\alpha $-dependent
term in Eq.~(\ref{pi00}), because by approaching the imaginary
pole ${\bf k}^2\to-m^2$,
the explicit factor that apparently ensures gauge independence
gets cancelled by a linear singularity in the momentum integral.
Exactly the same phenomenon was encountered\cite{BKS} in the recalculation
of plasmon damping rates in general covariant gauges.
I have argued previously\cite{BKSC} that this behaviour just
reflects a singular, gauge dependent behaviour of the {\it residue} of the
propagator rather than an actual gauge dependence of the pole
determining the dispersion laws. Indeed, introducing an (unphysical)
cut-off again moves the gauge dependence seemingly afflicting
the pole position into the residue, while the correction to the
pole position becomes independent of this infrared regularization.
Alternatively, the gauge dependent contributions can be avoided
altogether by a quantization procedure that keeps unphysical modes
unthermalized\cite{PVL}.
The gauge independent correction term then reads\cite{I}
\begin{equation}
\delta\equiv
{\delta m^2_{\rm el.}\0m^2_{0\rm el.}}
=gN\sqrt{\frac6{2N+N_f}}\frac1{2\pi}\left(
\ln\frac{m_{\rm el.}}{m_{\rm magn.}}
+\ln2-\frac12\right)+O(g^2).
\label{d}\end{equation}

In pure SU(2) lattice gauge theory high-statistics
results on Debye screening have rather recently been obtained\cite{ILMPR}
at temperatures well above the critical temperature, finding
a positive excess in the screening mass squared of $\d=+0.30(9)$.
Inserting the parameters of this lattice calculation as well as
an older lattice result for the magnetic screening mass\cite{MM},
(\ref{d}) yields $\d\approx +0.5$, which comes remarkably close
in view of $g\approx1$. 
(The renormalized value of $g$ used in this calculation appears to be
in good agreement with independent lattice calculations of the SU(2) pressure
when the latter is equated to the perturbative result.\cite{Kr})

In conclusion,
by defining the Debye mass through the relevant pole of the static
gluon propagator 
rather than the
(gauge-dependent) zero-momentum limit of the gluon self-energy,
a gauge-independent result at next-to-leading order has been
derived after identifying and resumming the relevant hard-thermal-loop
contributions. The location of the pole turns out to be logarithmically
sensitive to the nonperturbative magnetic screening mass, but the
coefficient of the corresponding logarithm can be calculated perturbatively.
The latter is in fact related\cite{Ip} to a logarithmic divergence
$\sim gm_{\rm el.}\ln(m_{\rm el.}r)$
encountered by Nadkarni\cite{N1} when trying to extract corrections
to the non-Abelian Debye mass from the correlation function of
two Polyakov loops. 

A similar logarithmic sensitivity to the scale $g^2T$ has been
encountered in the calculations of plasmon and fermion
damping\cite{LS}, which even appears in the Abelian case where
there is no magnetic screening mass. In the case of the Debye mass,
however, the origin of the
logarithmic term is genuinely non-Abelian. Another difference
which is important in view of the discussions\cite{Smilga}
surrounding the calculations on damping is that the position of
the pole defining the Debye mass is on the imaginary axis of $k$,
and it stays there when the corrections are included. This makes
it rather unnatural to attempt anything else than a self-consistent
procedure. As concerns the gauge dependences which in both
calculations require regularization of mass-shell singularities
(unless unphysical modes are frozen\cite{PVL}), it should
be kept in mind that they occur in the sublogarithmic terms
which are strictly speaking beyond the reach of perturbation
theory, at least in the non-Abelian case. It is
therefore left open whether the ticklish
``mass-shell singularities'' caused
by the massless unphysical modes in covariant gauges might not
disappear in a calculation which is complete down to
and including the order $k\sim g^2T$.

\vskip 14pt

I would like to thank R. Baier, J. Kapusta, U. Kraemmer,
R. Pisarski, K. Redlich, H. Schulz,
and A. Smilga for useful discussions, and the organizers of the
3rd Workshop on Thermal Field Theories, in particular R. Kobes and
G. Kunstatter, for this most enjoyable meeting.

\newpage

\end{document}